\documentclass[pra,twocolumn,preprintnumbers,amsmath,amssymb]{revtex4}
\usepackage{graphicx}
\usepackage{hyperref}

\begin{document}

\title{Nonperturbative thermodynamics of an interacting Bose gas}
\author{S. Floerchinger}
\author{C. Wetterich}
\affiliation{Institut f\"{u}r Theoretische Physik\\Universit\"at Heidelberg\\Philosophenweg 16, D-69120 Heidelberg, Germany}

\begin{abstract}We discuss the thermodynamics of a non-relativistic gas of bosons with a local repulsive interaction. In particular, we compute the temperature and density dependence of pressure, energy and entropy-density, superfluid and condensate-fraction, correlation length, specific heat, isothermal and adiabatic compressibility and various sound velocities. The $T\to0$ limit approaches the quantum phase transition, while the universal critical behavior of a classical second order phase transition in the O(2) universality class determines the region around the critical temperature. Our non-perturbative flow equations based on exact functional renormalization cover all regions in the phase diagram. 
\end{abstract}


\maketitle

\section{Introduction}
A gas of non-relativistic bosons with a repulsive pointlike interaction is one of the simplest interacting statistical systems. Since the first experimental realization \cite{FirstBEC} of Bose-Einstein condensation (BEC) \cite{BoseEinstein} with ultracold gases of bosonic atoms, important experimental advances  have been achieved, for reviews see \cite{ReviewsBECExperiment, PitaevskiiStringari}. Thermodynamic observables like the specific heat \cite{Oberthalercv} or properties of the phase transition like the critical exponent $\nu$ \cite{Donner} have been measured in harmonic traps. Still, the theoretical description of these apparently simple systems is far from being complete. This is partly due to the failure of perturbation theory for several interesting regions in the phase diagram: the quantum critical behavior in the limit of zero temperature \cite{BeliaevGavoretNepomnyashchii} or the critical behavior near the second order phase transition between the disordered and the BEC-phase. Also the generic infrared behavior in the BEC-phase, which is associated to a Goldstone boson, is non-perturbative \cite{InfraredGoldstone}. This issue becomes particularly important in two- or one-dimensional systems.

In this paper we employ non-perturbative flow equations for an investigation of the thermodynamic quantities for interacting bosons in three dimensions. In this approach the complications of the long-distance physics can be isolated and solved by a stepwise integration of the fluctuation effects. Already in a simple truncation we obtain a rather complete picture of the temperature and density dependence of many thermodynamic observables. In turn, if some of these observables can be measured precisely, and if the finite size effects due to the trap are properly taken into account, one may use such observables as compressibility or sound velocity as a precise ``thermometer'' of the system. Simultaneous knowledge of two independent thermodynamic observables will be sufficient to determine the thermodynamic parameters -- the temperature $T$ and the density $n$.

Furthermore, we study the dependence of the observables on the interaction strength. If the interaction strength can be varied experimentally, for example by its dependence on a magnetic fields, many aspects of our computations can be tested by observation. In particular, we have computed several non-analytic features near the phase transition. While the critical exponents and amplitude ratios in the immediate vicinity of the critical temperature $T_c$ are given by universal laws, the amplitudes themselves as well as the approach to criticality and the location and size of the critical region in the phase diagram are non-universal. The dependence of these quantities on the interaction strength may become an important test for non-perturbative methods. 

\section{Method}
\label{sectIntroduction}
The thermodynamic properties of a statistical system in thermal and chemical equilibrium, are described by the grand canonical partition function
\begin{equation}
Z=\text{Tr} e^{-\beta(H-\mu N)}=e^{-\beta \Omega_G}.
\end{equation}
The trace operation includes a summation over all possible states of the system including varying particle number $N$. We use $\beta=1/T$ with units where $\hbar=k_B=1$. The object $\Omega_G$ is the thermodynamic potential of the grand canonical partition function. It has the differential
\begin{equation}
d\Omega_G=-S\,dT-N\, d\mu-p\,dV.
\end{equation}
The partition function has a functional integral representation
\begin{equation}
Z=\int D\phi e^{-S[\phi]},
\label{eq:Zfunctint}
\end{equation}
where $\int D \varphi=\prod_{\{\tau,\vec x\}}\int d\varphi(\tau,\vec x)$ sums over all possible field configurations $\varphi(\tau,\vec x)$. The variable $\tau$ is a periodic euclidean time coordinate in the range $\tau\in(0,\beta)$ and $\vec x$ denotes the usual position coordinate. 

The microscopic action $S[\varphi]$ includes the Hamiltonian and the chemical potential. For nonrelativistic bosons with a pointlike interaction it is given by
\begin{equation}
S[\varphi]=\int_{\tau,\vec x}\left\{\varphi^*(\partial_\tau-\Delta-\mu)\varphi+\frac{1}{2}\lambda(\varphi^*\varphi)^2\right\}.
\label{eq:microscopicaction}
\end{equation}
We use energy units where $2M=1$, with $M$ the mass of the atoms. Apart from the chemical potential and the kinetic energy we include a repulsive pointlike interaction. The interaction strength $\lambda\geq 0$ determines the scattering length $a$. 
For $\lambda=0$ the functional integral in Eq. \eqref{eq:Zfunctint} is Gaussian and can be done analytically, however for $\lambda>0$ this is not possible any more. For small $\lambda$, or more precisely small $an^{1/3}$, one might try to make a perturbative analysis by expanding in $\lambda$, but this often leads to infrared problems. Furthermore, we also want to cover the case of strong interactions.

The method we use to determine the partition function in Eq. \eqref{eq:Zfunctint} is of a different kind. We do not perform the functional integral in one step. Instead, we integrate out fluctuations with large momenta first and fluctuations with small momenta later. From a physical point of view it is not reasonable to include the effect of fluctuations up to an infinetly large momentum scale. A microscopic action as in Eq. \eqref{eq:microscopicaction} is an effective description of the physics at some momentum scale $\Lambda$ (somewhat smaller than the inverse Bohr radius $a_0^{-1}$). At larger momentum or for larger energies, the physics is modified. For example, one might have to include photons and electrons into the description and at even larger energies quarks and gluons. All this is not visible anymore in the ``microscopic action'' \eqref{eq:microscopicaction} -- the fluctuations with momenta $q^2<\Lambda^2$ have already been ``integrated out''. Starting from Eq. \eqref{eq:microscopicaction}, the computations of thermodynamics should only include fluctuations with momenta smaller than the ``ultraviolet cutoff'' $\Lambda$. 

On the other hand, any real system or any given physical observation has also an effective ``infrared cutoff'' scale $k_\text{ph}$, such that only fluctuations in the range $k_\text{ph}^2<q^2<\Lambda^2$ matter. It makes no sense to include fluctuations with a wavelength larger than the size of some macroscopic probe. Alternatively, $k_\text{ph}$ may also be set by an inverse characteristic wavelength of some measurement device. At the scale $k_\text{ph}$ we would like to dispose of an object similar to the action \eqref{eq:microscopicaction}, with the difference that now all fluctuations with $q^2>k_\text{ph}^2$ instead of $q^2>\Lambda^2$ are integrated out. Such an ``average action'' \cite{Wetterich:1990an} $\Gamma_{k_\text{ph}}$ averages out all structures with wavelength smaller then $k_\text{ph}^{-1}$ and only retains the ``macroscopic'' information about physics at the momentum scale $k_\text{ph}$. It will constitute a Landau-type theory for the macroscopic physics. Thermodynamic quantities can be easily derived from $\Gamma_{k_\text{ph}}$, since no further fluctuations need to be taken into account. 

The basic challenge of a computation of thermodynamics from given microscopic laws is to build a bridge from the microscopic action $S=\Gamma_\Lambda$ to the average action $\Gamma_{k_\text{ph}}$. This can be a difficult task, since $\Gamma_{k_\text{ph}}$ may be rather different from $\Gamma_\Lambda$. Typically, a large part of the precise microscopic information is lost in the averaging procedure. On the other hand, new phenomena due to collective effects may appear in $\Gamma_{\text{ph}}$, which are not directly visible in $\Gamma_\Lambda$. The basic idea of our approach is to make the infrared cutoff scale $k$ variable. The resulting ``{\itshape flowing action}'' or scale dependent average action $\Gamma_k$ interpolates continuously between the microscopic action for $k=\Lambda$ and the macroscopic action $\Gamma_{k_\text{ph}}$. Lowering $k$ from $\Lambda$ to $k_\text{ph}$ the fluctuation effects are included stepwise. In this sense it realizes the Wilsonian idea of renormalization, even though the implementation of a sliding infrared cutoff leads to several important conceptual and technical differences as compared to the sliding ultraviolet cutoff investigated in the first approaches to functional renormalization \cite{FunctionalRenormalization}. 

The dependence of the flowing action $\Gamma_k$ on the infrared cutoff scale $k$ obeys an exact functional differential equation \cite{Wetterich:1992yh}. It can be solved approximately by a truncation of the most general functional form of the flowing action. Such truncations do not have to rely on the expansion in some small parameter as the interaction strength and can describe physical phenomena that are ``non-perturbative''. 

The average potential $U_k(\bar \rho)$ obtains from $\Gamma_k[\bar \varphi]$ by using for the argument a homogeneous (and $\tau$-independent) field $\bar \varphi$, with $\bar \rho=\bar \varphi^*\bar\varphi$. An evaluation of the potential at its minimum, $U_\text{min}$, yields directly the grand canonical partition function. 
\begin{equation}
Z=e^{-\beta \Omega_G}=e^{\beta V U_\text{min}},
\end{equation}
with
\begin{equation}
U_\text{min}=U_{k=0}(\bar \rho_0)\quad \text{with}\quad \frac{\partial}{\partial \bar \rho}U_{k=0}(\bar \rho){\bigg |}_{\bar \rho=\bar \rho_0}=0.
\end{equation}
The expectation value or order parameter $\bar \rho_0$ can be associated with the condensate density. The average potential $U_k(\bar \rho)$ obeys the exact flow equation
\begin{equation}
k\frac{\partial}{\partial k}U_k(\bar \rho)=\frac{1}{2}\int_{q_0}\int_{\vec q}\frac{(P_{11}+P_{22}+2 R_k) k\partial_k R_k}{(P_{11}+R_k)(P_{22}+R_k)+P_{12}^2}.
\end{equation}
We use the Matsubara formalism with
\begin{eqnarray}
\int_{q_0}=T\sum_n,\quad q_0=2\pi n T,\quad n\in \mathrm{N},\quad
\int_{\vec q}=\int \frac{d^3 q}{(2\pi)^3}.
\end{eqnarray}

In our truncation we approximate the momentum dependence of the inverse propagator for the radial and angular (Goldstone) mode by
\begin{eqnarray}
\nonumber
P_{11}(q) &=& \bar A \vec q^2+\bar V q_0^2+U^\prime+2\bar \rho U^{\prime\prime}\\
\nonumber
P_{22}(q) &=& \bar A \vec q^2+\bar V q_0^2+U^\prime\\
P_{12}(q) &=& \bar S q_0.
\end{eqnarray}
The flow equations for the quantities $\bar A$, $\bar S$, and $\bar V$ can be found in ref. \cite{FWBEC}. 

The effective potential $U$ is related to the pressure by
\begin{equation}
p(T,\mu)=-U_\text{min}(T,\mu){\big |}_{k=k_\text{ph}}
\end{equation}
which has the differential
\begin{equation}
dp=s\,dT+n\,d\mu.
\label{eq:differentialp}
\end{equation}
Here we use $s=S/V$ for the entropy density and $n=N/V$ for the particle density. The formal infinite volume limit corresponds to $k_\text{ph}=0$. We can use our method to determine many thermodynamic quantities. Derivatives of $U$ with respect to $T$ and and $\mu$ are taken numerically by solving the flow equation for close enough values of $T$ and $\mu$. The numerical effort is reduced and the accuracy increased by using an additional flow equation for
\begin{equation}
n_k=-\frac{\partial}{\partial \mu} U_k {\big |}_{\bar \rho=\bar \rho_0(k)},
\end{equation}
with $n=n_{k_\text{ph}}$. The details of our method as well as explicit expressions for the flow equations can be found in \cite{FWBEC}. The approximation scheme we use in this paper is basically the same as the one used there. Since we use an infrared cutoff only for momenta but not for frequencies, the correct ultraviolet convergence for the sum of the Matsubara frequencies is not automatically obeyed for the flow equations. We have checked that all thermodynamic quantities discussed in this paper show a satisfactory convergence of the Matsubara sum, except for the pressure. In the flow equation for $p_k$ we set the frequency coefficients to their microscopic values $\bar S=1$, $\bar V=0$ for very large Matsubara frequencies $|q_0|>\Lambda_\text{UV}^2$. 

For bosons with a pointlike repulsive interaction we found in \cite{FWBEC} that the scattering length is bounded by the ultraviolet scale $a<3\pi/(4\Lambda)$. This is an effect due to quantum fluctuations similar to the ``triviality bound'' for the Higgs scalar in the standard model of elementary particle physics. For a given value of the dimensionless combination $an^{1/3}$ we cannot choose $\Lambda/n^{1/3}$ larger then $3\pi/(4 an^{1/3})$. For our numerical calculations we use $\Lambda/n^{1/3}\approx 10$. Other momentum scales are set by the temperature and the chemical potential. The lowest nonzero Matsubara frequency gives the momentum scale $\Lambda_T^2=2\pi T$. For a Bose gas with $a=0$ one has $T_c/n^{2/3}\approx 6.625$ such that $\Lambda_{T_c}/n^{1/3}\approx 6.45$. The momentum scale associated to the chemical potential is $\Lambda_\mu^2=\mu$. For small temperatures and scattering length one finds $\mu\approx 8\pi a n$ and thus $\Lambda_\mu/n^{1/3}\approx \sqrt{8\pi a n^{1/3}}$.

We finally note that the thermodynamic relations for intensive quantities can only involve dimensionless ratios. We may set the unit of momentum by $n^{1/3}$. The thermodynamic variables are then $T/n^{2/3}$ and $\mu/n^{2/3}$. The thermodynamic relations will depend on the strength of the repulsive interaction $\lambda$ or the scattering length $a$, and therefore on a ``concentration'' type parameter $a n^{1/3}$.

\section{Density, superfluid density, condensate and correlation length}
\begin{figure}
\includegraphics[width=\linewidth]{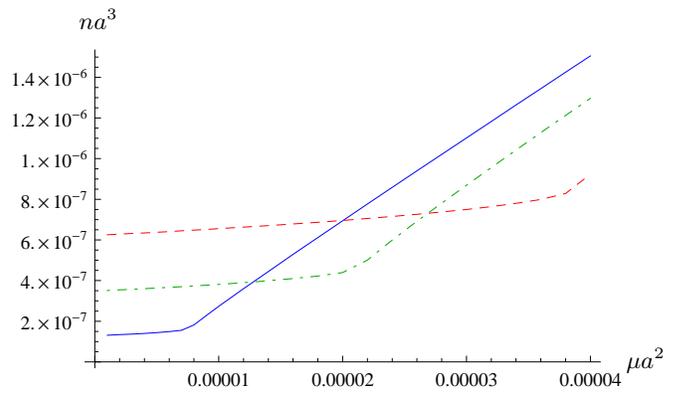}
\caption{(Color online) Density in units of the scattering length $n a^3$ as a function of the (rescaled) chemical potential $\mu a^2$. We choose for the temperatures $T a^2=2\cdot 10^{-4}$ (solid curve), $T a^2=4 \cdot 10^{-4}$ (dashed-dotted curve) and $T a^2= 6 \cdot 10^{-4}$ (dashed curve). For all three curves we use $a\Lambda=0.1$.}
\label{fig:densityofmu}
\end{figure}

Let us start our discussion of the thermodynamic properties with the density. In the grand canonical formalism it is obtained by taking the derivative of the thermodynamic potential with respect to $\mu$
\begin{equation}
n=-\frac{1}{V}\frac{\partial}{\partial \mu}\Omega_G = \frac{\partial p}{\partial \mu}{\big |}_T.
\end{equation}
We could compute the $\mu$-derivative of $p$ numerically by solving the flow equation for U with neighboring values of $\mu$. In \cite{FWBEC} we also describe another method which employs a flow equation directly for $n$. Since we often express dimensionful quantities in units of the interparticle distance $n^{-1/3}$, it is crucial to have an accurate value for the density $n$. Comparison of the numerical evaluation and the solution of a separate flow equation for $n$ shows higher precision for the latter method and we will therefore employ the flow equation. We plot in Fig. \ref{fig:densityofmu} the density in units of the scattering length, $na^3$, as a function of the dimensionless combination $\mu a^2$. 

\begin{figure}
\includegraphics[width=\linewidth]{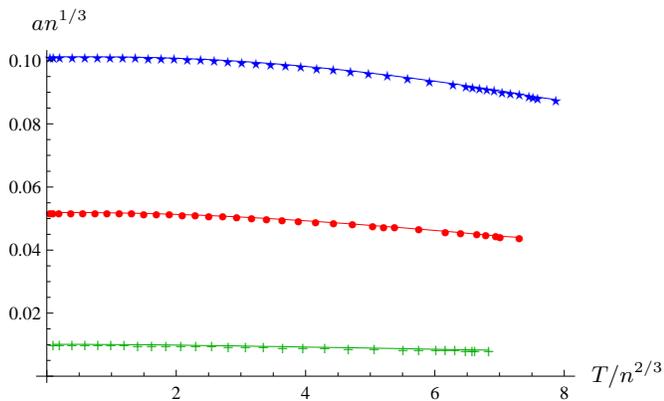}
\caption{(Color online) Concentration $c=an^{1/3}$ as a function of temperature $T/(n^{2/3})$ for the three cases investigated in this paper. Case I corresponds to $an^{1/3}\approx 0.01$ (crosses), case II to $an^{1/3}\approx 0.05$ (dots) and case III has $an^{1/3}\approx 0.01$ (stars).}
\label{fig:an13}
\end{figure}

For a comparison with experimentally accessible quantities we have to replace the interaction parameter $\lambda$ in the microscopic action \eqref{eq:microscopicaction} by a scattering length $a$ which is a macroscopic quantity. For this purpose we start the flow at the UV-scale $\Lambda_\text{UV}$ with a given $\lambda$, and then compute the scattering length in vacuum ($T=n=0$) by following the flow to $k=0$ \cite{FWBEC}. This is a standard procedure in quantum field theory, where a ``bare coupling'' ($\lambda$) is replaced by a renormalized coupling ($a$). For an investigation of the role of the strength of the interaction we may consider different values of the ``concentration'' $c=an^{1/3}$ or of the product $\mu a^2$. While the concentration is easier to access for observation, it is also numerically more demanding since for every value of the parameters one has to tune $\mu$ in order to obtain the appropriate density. For this reason we rather present results for three values of $\mu a^2$, i.~e. $\mu a^2= 2.6\times 10^{-5}$ (case I), $\mu a^2=0.0040$ (case II) and $\mu a^2=0.044$ (case III). The prize for the numerical simplicity is a week temperature dependence of the concentration $c=a n^{1/3}$ for the three different cases, as shown in Fig. \ref{fig:an13}.
Here and in the following figures case I, which corresponds to $an^{1/3}\approx 0.01$, is represented by the little crosses, case II with $an^{1/3}\approx 0.05$ by the dots and case III with $an^{1/3}\approx 0.1$ by the stars. It is well known that the critical temperature depends on the concentration $c=an^{1/3}$. From our calculation we find $T_c/(n^{2/3})=6.74$ with $c=0.0083$ at $T=T_c$ in case I, $T_c/(n^{2/3})=7.16$ with $c=0.044$ at $T=T_c$ in case II and finally $T_c/(n^{2/3})=7.75$ with $c=0.088$ at $T=T_c$ in case III.

\begin{figure}
\includegraphics[width=\linewidth]{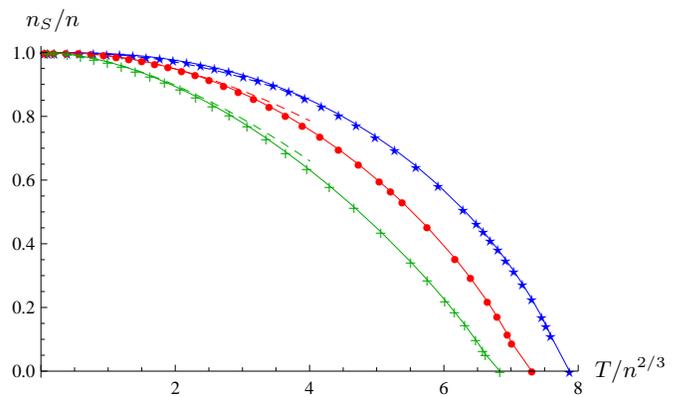}
\caption{(Color online) Superfluid fraction of the density $n_S/n$ as a function of the temperature $T/n^{2/3}$ for the cases I, II, and III. For small $T/n^{2/3}$ we also show the corresponding curves obtained in the Bogoliubov approximation (dashed lines).}
\label{fig:superfluidfraction}
\end{figure}

\begin{figure}
\includegraphics[width=\linewidth]{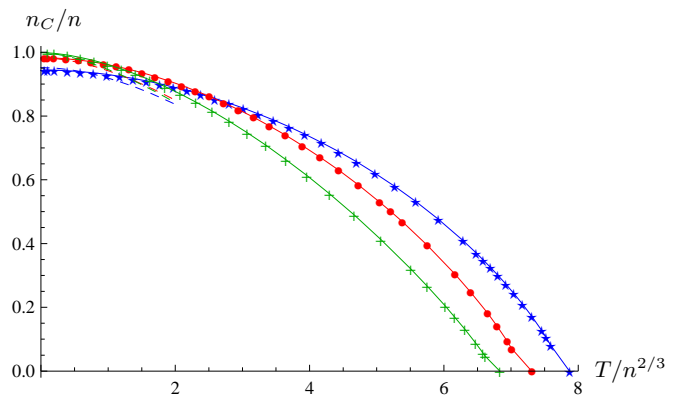}
\caption{(Color online) Condensate fraction of the density $n_C/n$ as a function of the temperature $T/n^{2/3}$ for the cases I, II, and III. For small $T/n^{2/3}$ we also show the corresponding curves obtained in the Bogoliubov approximation (dashed lines).}
\label{fig:condensatefraction}
\end{figure}

This values can are obtained by following the superfluid fraction of the density $n_S/n$, or equivalently the condensate part of the density $n_C/n$ as a function of temperature. For small temperatures $T\to0$ all of the density is superfluid, which is a consequence of Galilean symmetry. However, in contrast to the ideal gas, not all particles are in the condensate. For $T=0$ this condensate depletion is completely due to quantum fluctuations. With increasing temperature both the superfluid density and the condensate decrease and vanish eventually at the critical temperature $T=T_c$. That the melting of the condensate is continuous shows that the phase transition is of second order. We plot our results for the superfluid fraction in Fig. \ref{fig:superfluidfraction} and for the condensate in Fig. \ref{fig:condensatefraction}. For small temperatures, we also show the corresponding result obtained in the framework of Bogoliubov theory \cite{Bogoliubov} (dashed lines). This approximation assumes a gas of non-interacting quasiparticles (phonons) with dispersion relation
\begin{equation}
\epsilon(p)=\sqrt{2\lambda n \vec p^2+\vec p^4}.
\end{equation} 
It is is valid in the regime with small temperatures $T\ll T_c$ and small interaction strength $an^{1/3}\ll 1$. For a detailed discussion of Bogoliubov theory and the calculation of thermodynamic observables in this framework we refer to ref. \cite{PitaevskiiStringari}. Our curves for the superfluid fraction match the Bogoliubov result for temperatures $T/n^{2/3}\lesssim 1$ in all three cases I, II, and III. For larger temperatures there are deviations as expected. For the condensate density, there is already notable a deviation at small temperatures for case III with $an^{1/3}\approx 0.1$. This is also expected, since Bogoliubov theory gives only the first order contribution to the condensate depletion in a perturbative expansion for small $an^{1/3}$.  
For temperatures slightly smaller than the critical temperature $T_c$ one expects that the condensate density behaves like 
\begin{equation}
n_c(T)=B^2 \left(\frac{T_c-T}{T_c}\right)^{2\beta}
\label{eq:scalingnc}
\end{equation}
with $\beta=0.3485$ the critical exponent of the three-dimensional XY-universality class \cite{Pelissetto:2000ek}. Indeed, the condensate density is given by $n_C=\bar\phi_0^*\bar\phi_0$
where $\bar \phi_0$ is the expectation value of the boson field which serves as an order parameter in close analogy to e.~g. the magnetization $\vec M$ in a ferromagnet. Eq. \eqref{eq:scalingnc} is compatible with our findings, although our numerical resolution does not allow for a precise determination of the exponent $\beta$. 

\begin{figure}
\includegraphics[width=\linewidth]{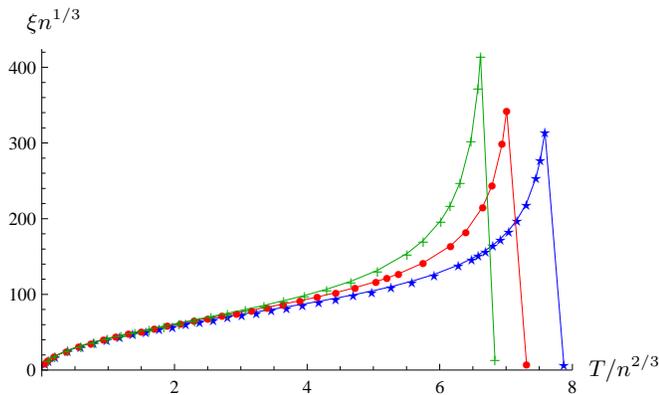}
\caption{(Color online) Correlation length $\xi_R n^{1/3}$ for $T<T_c$ and $\xi n^{1/3}$ for $T>T_c$ as a function of the temperature $T/n^{2/3}$ for the cases I, II, and III.}
\label{fig:correlationlength}
\end{figure}

With our method we can also calculate the correlation length $\xi$. For temperatures $T<T_c$ one distinguishes between the Goldstone correlation length $\xi_G$ and the radial correlation length $\xi_R$. While the former is infinite, $\xi_G^{-1}=0$, the latter is finite for $T<T_c$. It is also known as the ``healing length'', given by 
\begin{equation}
\xi_R^{-2}=2\lambda \rho_0=2\frac{1}{\bar A}\frac{\partial^2 U}{\partial \bar \rho}\,\bar \rho_0
\end{equation} 
and diverges only close to the phase transition. In the symmetric regime for $T>T_c$ there is only one correlation length $\xi^{-1}=m=\frac{1}{\bar A}\frac{\partial U}{\partial \bar \rho}$, which also diverges for $T\to T_c$. From the theory of critical phenomena one expects close to $T_c$ the behavior
\begin{eqnarray}
\nonumber
\xi_R &=& f_R^- \left(\frac{T_c-T}{T_c}\right)^{-\nu} \quad \text{for}\quad T<T_c\\
\xi &=& f^+ \left(\frac{T-T_c}{T_c}\right)^{-\nu} \quad \text{for}\quad T<T_c.
\end{eqnarray}
The critical exponent $\nu=0.6716$ \cite{Pelissetto:2000ek} is again the one of the three-dimensional XY- or O(2) universality class. We plot our result for the correlation length in units of the interparticle distance $\xi_R n^{1/3}$ for $T<T_c$ and $\xi n^{1/3}$ for $T>T_c$ as a function of the temperature $T/n^{2/3}$ in Fig. \ref{fig:correlationlength}.

\section{Entropy density, energy density, and specific heat}
\begin{figure}
\includegraphics[width=\linewidth]{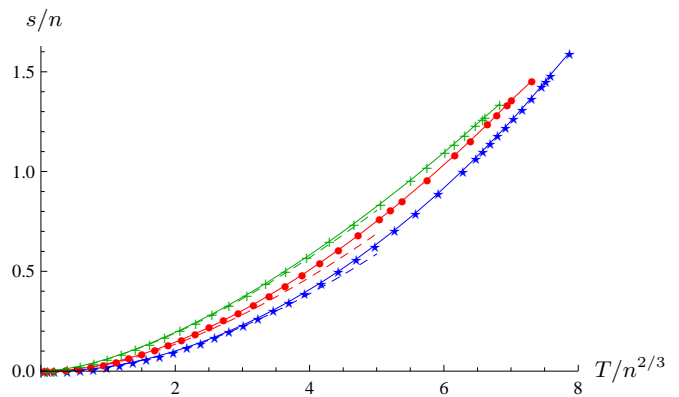}
\caption{(Color online) Entropy per particle $s/n$ as a function of the dimensionless temperature $T/n^{2/3}$ for the cases I, II, and III. For $T/n^{2/3}<5$ we also plot the results obtained within the Bogoliubov approximation (dashed lines).}
\label{fig:entropypp}
\end{figure}

The next thermodynamic quantity we investigate is the entropy density $s$ and the entropy per particle $s/n$. We can obtain the entropy as
\begin{equation}
s=\frac{\partial p}{\partial T}{\big |}_\mu.
\end{equation}
We compute the temperature derivative by numerical differentiation, using flows with neighboring values of $T$ and show the result in Fig. \ref{fig:entropypp}. For small temperatures our result coincides with the entropy of free quasiparticles in the Bogoliubov approximation (dashed lines in Fig. \ref{fig:entropypp}). 
As it should be, the entropy per particle increases with the temperature. For small temperatures, the slope of this increase is smaller for larger concentration $c$.
We may consider a change of the volume $V$ by a change in the trap, for example increasing $V$ by making the trap shallower. If no particles are lost, $N=\text{const.}$, an increase of $V$ corresponds to a decrease of $n$. Furthermore, if the change is adiabatic, the entropy and the entropy per particle, $s/n$, remains constant. For constant $s/n$ the ratio $T/n^{2/3}=\gamma$ is fixed, as given by Fig. \ref{fig:entropypp} for given $s/n$. An adiabatic increase of the volume therefore induces a lowering of the temperature, $T=\gamma n^{2/3}\sim V^{-2/3}$. This allows for a continuous reversible variation of the temperature by means of a variation of the trap parameters. By an adiabatic increase of $a$ (at fixed $N$) one can increase the ratio $T/n^{2/3}$. This can be realized by a variation of a magnetic field, which may therefore be used to explore the phase transition and the region of $T/n^{2/3}$ around the critical temperature. 

\begin{figure}
\includegraphics[width=\linewidth]{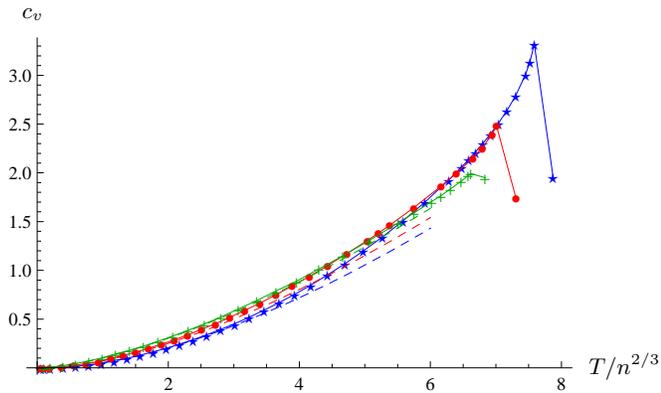}
\caption{(Color online) Specific heat per particle $c_v$ as a function of the dimensionless temperature $T/n^{2/3}$. The dashed lines show the Bogoliubov result for $c_v$ which coincides with our findings for small temperature. However, the characteristic cusp behavior cannot be seen in a mean-field theory.}
\label{fig:specificheatpp}
\end{figure}

From the entropy density $s$ we infer the specific heat per particle,
\begin{equation}
c_v=\frac{T}{n}\frac{\partial s}{\partial T}{\bigg |}_n,
\end{equation}
as the temperature derivative of the entropy density at constant particle density.
Using the Jacobian, we can write
\begin{equation}
\frac{\partial s}{\partial T}{\big |}_n = \frac{\partial(s,n)}{\partial(T,n)}=\frac{\partial(s,n)}{\partial(T,\mu)}\frac{\partial(T,\mu)}{\partial(T,n)}.
\end{equation}
For the specific heat this gives
\begin{equation}
c_v=\frac{T}{n}\left(\frac{\partial s}{\partial T}{\big |}_\mu -\frac{\partial s}{\partial \mu}{\big |}_T\frac{\partial n}{\partial T}{\big |}_\mu \left(\frac{\partial n}{\partial \mu}{\big |}_T\right)^{-1} \right).
\end{equation}
Our result for the specific heat per particle is shown for different scattering lengths in Fig. \ref{fig:specificheatpp}.
While this quantity is positive in the whole range of investigated temperatures, it is interesting to observe the cusp at the critical temperature $T_c$ which is characteristic for a second order phase transition. This behavior cannot be seen in a mean-field approximation, where fluctuations are taken into account only to second order in the fields. Only for small temperatures, our curve is close to the Bogoliubov approximation, shown by the dashed lines in Fig. \ref{fig:specificheatpp}. 

In fact, close to $T_c$ the specific heat is expected to behave like
\begin{eqnarray}
\nonumber
c_v &\approx&  b_1-b_2^- \left(\frac{T_c-T}{T_c}\right)^{-\alpha} \quad \text{for} \quad T<T_c,\\
c_v &\approx&  b_1-b_2^+ \left(\frac{T-T_c}{T_c}\right)^{-\alpha} \quad \text{for} \quad T>T_c,
\end{eqnarray}
with the universal critical exponent $\alpha$ of the $3$-dimensional $XY$ universality class, $\alpha=-0.0146(8)$ \cite{Pelissetto:2000ek}. The critical region, where the law $c_v~\sim |T-T_c|^{-\alpha}$ holds, may be quite small.
Our numerical differentiation procedure cannot resolve the details of the cusp.  

\begin{figure}
\includegraphics[width=\linewidth]{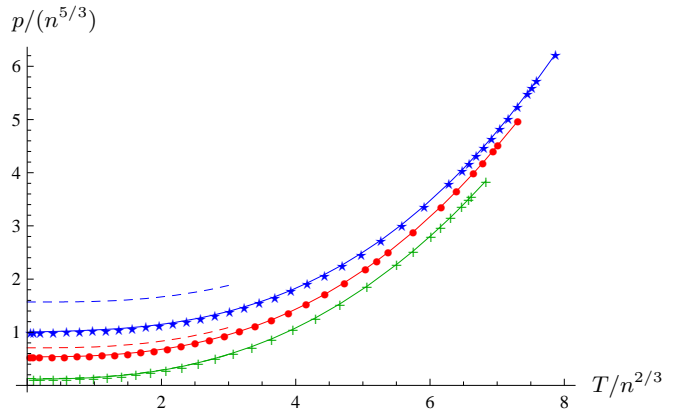}
\caption{(Color online) Pressure in units of the density $p/n^{5/3}$ as a function of temperature $T/n^{2/3}$ for the cases I (crosses), II (dots), and III (stars). We also show the curves obtained in the Bogoliubov approximation for small temperatures (dashed lines).}
\label{fig:pressure}
\end{figure}

\begin{figure}
\includegraphics[width=\linewidth]{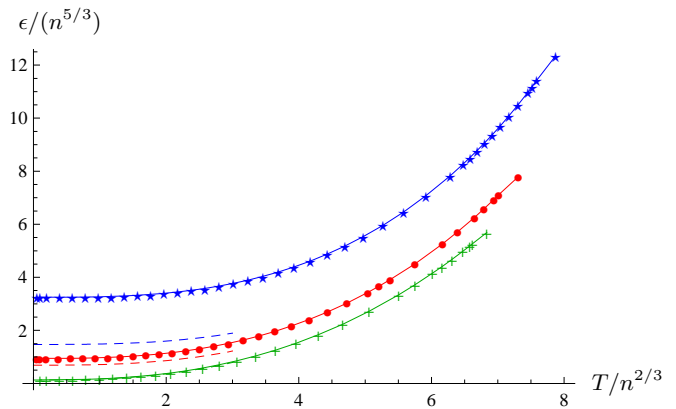}
\caption{(Color online) Energy per particle $\epsilon/n^{5/3}$ as a function of temperature $T/n^{2/3}$ for the cases I (crosses), II (dots), and III (stars). We also show the curves obtained in the Bogoliubov approximation for small temperatures (dashed lines).}
\label{fig:energy}
\end{figure}

In the grand canonical formalism, the energy density $\epsilon$ is obtained as
\begin{equation}
\epsilon = -p +Ts+\mu n.
\end{equation}
(The total energy density density in the gravitational context reads $n M c^2+\epsilon$.) 
We plot $p/(n^{5/3})$ as a function of temperature in Fig. \ref{fig:pressure} and the energy density $\epsilon/(n^{5/3})$ is plotted in Fig. \ref{fig:energy}.
We have normalized the pressure such that it vanishes for $T=\mu=0$. Technically we subtract from the flow equation of the pressure the corresponding expression in the limit $T=\mu=0$. This procedure has to be handled with care and leads to an uncertainty in the offset of the pressure, i.~e. the part that is independent of $T/n^{2/3}$ and $\mu/n^{2/3}$. 

For zero temperature, the pressure is completely due to the repulsive interaction between the particles. For nonzero temperature, the pressure is increased by the thermal kinetic energy, of course.
For the energy and the pressure we find some deviations from the Bogoliubov result already for small temperatures in cases II and III. These deviations may be partly due to the uncertainty in the normalization process described above. For weak interactions $an^{1/3}=0.01$ as in case I, the Bogoliubov prediction coincides with our result. 

\section{Compressibility}
The isothermal compressibility is defined as the relative volume change at fixed temperature $T$ and particle number $N$ when some pressure is applied
\begin{equation}
\kappa_T=-\frac{1}{V}\frac{\partial V}{\partial p}{\big |}_{T,N}=\frac{1}{n}\frac{\partial n}{\partial p}{\big |}_T.
\label{eq:isothermalkomp}
\end{equation}
Very similar, the adiabatic compressibility is
\begin{equation}
\kappa_S=-\frac{1}{V}\frac{\partial V}{\partial p}{\big |}_{S,N}=\frac{1}{n}\frac{\partial n}{\partial p}{\big |}_{s/n}
\end{equation}
where now the entropy $S$ and the particle number $N$ are fixed. Let us first concentrate on the isothermal compressibility $\kappa_T$. To evaluate it in the grand canonical formalism, we have to change variables to $T$ and $\mu$. 
With $\partial p/\partial \mu{\big |}_{n,T}=n$ and $\partial p/\partial n{\big |}_T=n \partial \mu/\partial n{\big |}_T$ one obtains
\begin{equation}
\kappa_T=\frac{1}{n^2}\frac{\partial n}{\partial \mu}{\big |}_{T}.
\end{equation}
This expression can be directly evaluated in our formalism by numerical differentiation with respect to $\mu$.  

\begin{figure}
\includegraphics[width=\linewidth]{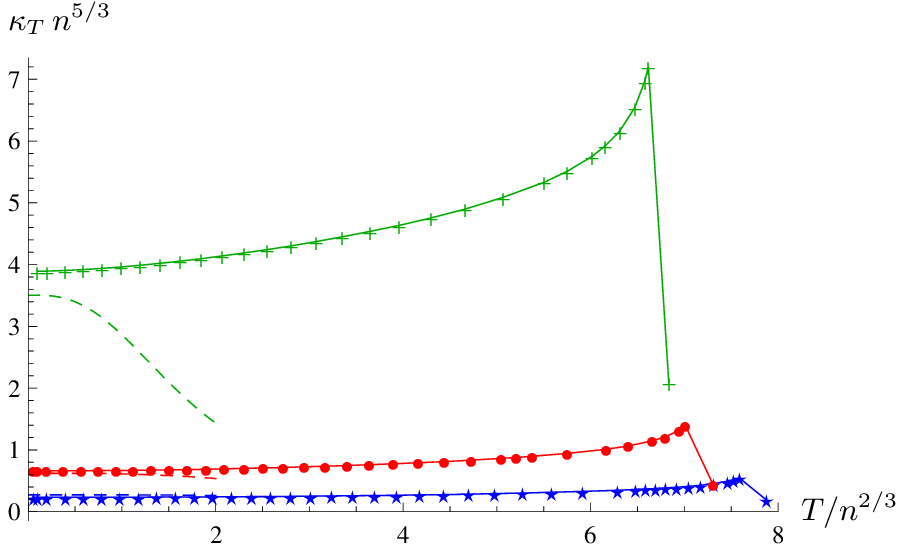}
\caption{(Color online) Isothermal compressibility $\kappa_T\, n^{5/3}$ as a function of temperature $T/n^{2/3}$ for the cases I (crosses), II (dots), and III (stars). We also show the Bogoliubov result for small temperatures (dashed lines).}
\label{fig:Isothermalcompressibility}
\end{figure}

\begin{figure}
\includegraphics[width=\linewidth]{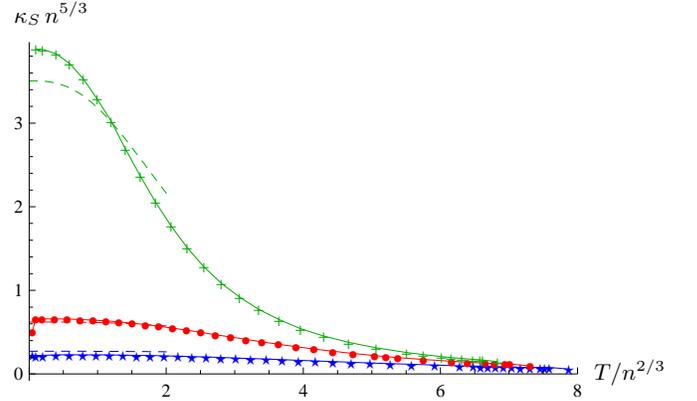}
\caption{(Color online) Adiabatic compressibility $\kappa_S\, n^{5/3}$ as a function of temperature $T/n^{2/3}$ for the cases I (crosses), II (dots), and III (stars). We also show the Bogoliubov result for small temperatures (dashed lines).}
\label{fig:Adiabaticcompressibility}
\end{figure}

The approach to the adiabatic compressibility is similar. Using again the Jacobian we have
\begin{eqnarray}
\nonumber
\kappa_S &=& \frac{1}{n}\frac{\partial n}{\partial p}{\big |}_{s/n}=\frac{1}{n}\frac{\partial(n,s/n)}{\partial(p,s/n)}\\
&=& \frac{1}{n}\frac{\partial (n,s/n)}{\partial (\mu,T)}\frac{\partial(\mu,T)}{\partial(p,s/n)}.
\end{eqnarray}
We need therefore
\begin{equation}
\frac{\partial(n,s/n)}{\partial(\mu,T)}=\frac{1}{n}\left(\frac{\partial n}{\partial \mu}{\big |}_{T}\frac{\partial s}{\partial T}{\big |}_{\mu}-\frac{\partial n}{\partial T}{\big |}_{\mu}\frac{\partial s}{\partial \mu}{\big |}_{T}\right)
\end{equation}
and also
\begin{eqnarray}
\nonumber
\frac{\partial(p,s/n)}{\partial(\mu,T)} &=& \frac{1}{n}{\bigg (}\frac{\partial p}{\partial \mu}{\big |}_{T}\frac{\partial s}{\partial T}{\big |}_\mu - \frac{\partial p}{\partial \mu}{\big |}_{T} \frac{s}{n}\frac{\partial n}{\partial T}{\big |}_\mu\\
&& -\frac{\partial p}{\partial T}{\big |}_{\mu} \frac{\partial s}{\partial \mu}{\big |}_T+\frac{\partial p}{\partial T}{\big |}_\mu \frac{s}{n} \frac{\partial n}{\partial \mu}{\big |}_T {\bigg )}\\
\nonumber
&=& \left(\frac{\partial s}{\partial T}{\big |}_\mu-2\frac{s}{n}\frac{\partial n}{\partial T}{\big |}_\mu+\frac{s^2}{n^2}\frac{\partial n}{\partial \mu}{\big |}_T\right).
\end{eqnarray}
In the last equations we used the Maxwell identity $\frac{\partial n}{\partial T}{\big |}_\mu=\frac{\partial s}{\partial \mu}{\big |}_T$. Combining this we find
\begin{equation}
\kappa_S=\frac{\left(\frac{\partial n}{\partial \mu}{\big |}_T \frac{\partial s}{\partial T}{\big |}_\mu-\left(\frac{\partial n}{\partial T}{\big |}_\mu\right)^2\right)}{\left(n^2 \frac{\partial s}{\partial T}{\big |}_\mu-2 s n \frac{\partial n}{\partial T}{\big |}_\mu+s^2 \frac{\partial n}{\partial \mu}{\big |}_T\right)}.
\end{equation}
Since $\partial s/\partial T{\big |}_\mu=(\partial^2 p/\partial T^2){\big |}_\mu$ we need to evaluate a second derivative numerically.
We plot the isothermal and the adiabatic compressibility in Figs. \ref{fig:Isothermalcompressibility} and \ref{fig:Adiabaticcompressibility}.

For the isothermal compressibility the temperature dependence is qualitatively different than in Bogoliubov theory already for small temperatures, while there seem to be only quantitative differences for the adiabatic compressibility. The perturbative calculation of the compressibility is difficult since it is diverging in the non-interacting limit $an^{1/3}\to 0$.

\section{Isothermal and adiabatic sound velocity}
\begin{figure}
\includegraphics[width=\linewidth]{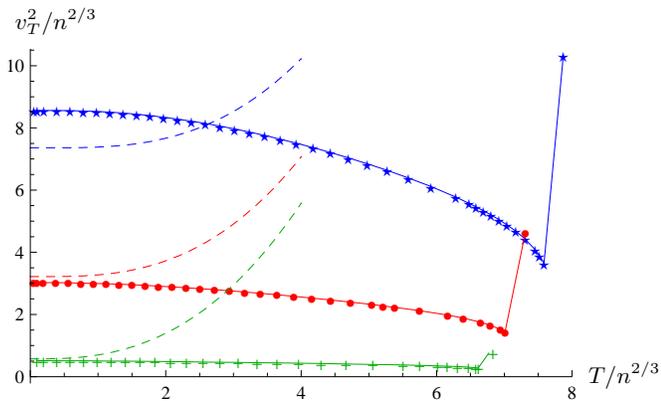}
\caption{(Color online) Isothermal velocity of sound as appropriate for single fluid $v_T^2/n^{2/3}=1/(\kappa_T\, n^{5/3})$ as a function of the dimensionless temperature $T/n^{2/3}$ for the cases I (crosses), II (dots), and III (stars). We also show the Bogoliubov result for small temperatures (dashed lines).}
\label{fig:SingleFluidisothermalsound}
\end{figure}

The sound velocity of a normal fluid under isothermal conditions, i.~e. for constant temperature $T$ is given by
\begin{equation}
v_T^2=\frac{1}{M}\frac{\partial p}{\partial n}{\big |}_T.
\label{eq:singlefluidisothermalsound}
\end{equation} 
We can obtain this directly from the isothermal compressibility
\begin{equation}
M v_T^2=(n \kappa_T)^{-1}
\end{equation}
as follows from Eq. \eqref{eq:isothermalkomp}. We plot our result for $v_T^2$ in Fig. \ref{fig:SingleFluidisothermalsound}, recalling our units $2M=1$ such that $v_T^2$ stands for $2Mv_T^2$.
This plot also covers the superfluid phase where the physical meaning of $v_T^2$ is partly lost. This comes since the sound propagation there has to be described by more complicated two-fluid hydrodynamics. In addition to the normal gas there is now also a superfluid fraction allowing for an additional oscillation mode. We will describe the consequences of this in the next section.

\begin{figure}
\includegraphics[width=\linewidth]{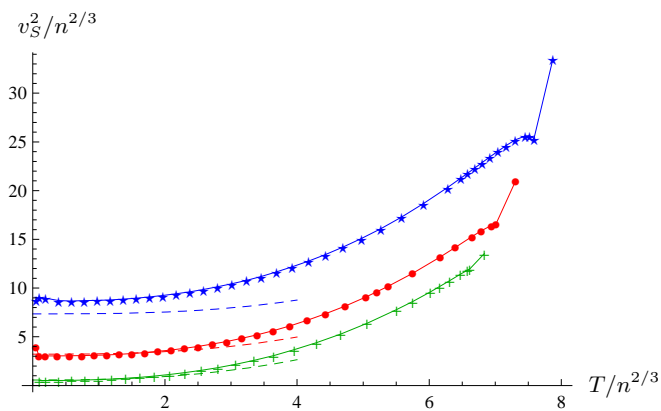}
\caption{(Color online) Adiabatic velocity of sound as appropriate for single fluid $v_S^2/n^{2/3}=1/(\kappa_S \,n^{5/3})$ as a function of the dimensionless temperature $T/n^{2/3}$ for the cases I (crosses), II (dots), and III (stars). We also show the Bogoliubov result for small temperatures (dashed lines).}
\label{fig:SingleFluidadiabaticsound}
\end{figure}

For most applications the adiabatic sound velocity is more important then the isothermal sound velocity. Keeping the entropy per particle fixed, we obtain
\begin{equation}
v_S^2=\frac{1}{M}\frac{\partial p}{\partial n}{\big |}_{s/n}
\label{eq:singlefluidadiabaticsound}
\end{equation}
and therefore
\begin{equation}
M v_S^2 = (n\kappa_S)^{-1}.
\end{equation}
Our numerical result is plotted in Fig. \ref{fig:SingleFluidadiabaticsound}.
Again the plot covers both the superfluid and the normal part, but only in the normal phase the object $v_S^2$ has its physical meaning as a sound velocity.

\section{First and second velocity of sound}
For temperatures $0<T<T_c$ there are two components of the gas: the superfluid and the normal part. It was shown by Landau \cite{Landau41} that this leads to two-fluid hydrodynamics with two distinct velocities of sound $c_{1/2}$ corresponding to different kinds of excitations. 

The main reason for the existence of two sound velocities is that the entropy flow is carried only be the normal component while the particle flow (or equivalently mass-flow) is carried by both the normal and the superfluid part. The continuity equation for the conserved particle number reads
\begin{equation}
\partial_t n+\vec \nabla \cdot \vec j =0,
\label{eq:particlecontinuity}
\end{equation}
where $\vec j=n_N \vec v_N+n_S \vec v_S$ is the (complete) particle number current and $\vec v_N$, $\vec v_S$ are the velocities of the normal ($n_N$) and superfluid ($n_S$) parts of the density, $n=n_N+n_S$. The conservation equation for the entropy reads 
\begin{equation}
\partial_t s + \vec \nabla \cdot (s \vec v_N)=0.
\label{eq:entropycontinuity}
\end{equation}
We work in linear order in an expansion in the velocities $\vec v_N$ and $\vec v_S$.
To close the set of hydrodynamic equations for small $\vec v_N$, $\vec v_S$ we need the equations for momentum conservation
\begin{equation}
M \partial_t \vec j+\vec \nabla p=0,
\label{eq:momentumconservation}
\end{equation}
and for the change in the superfluid velocity
\begin{equation}
M\partial_t \vec v_S + \vec \nabla \mu=0.
\label{eq:changesuperfluidvelocity}
\end{equation}
The last equation guarantees that the superfluid flow remains irrotational, $\vec \nabla \times \vec v_S=0$.

From the combination of Eq. \eqref{eq:particlecontinuity} and \eqref{eq:momentumconservation} one obtains
\begin{equation}
M \partial_t^2 n = \Delta p.
\label{eq:wave1}
\end{equation}
To linear order in $\vec v_S$ and $\vec v_N$ one infers from the combination of Eq. \eqref{eq:particlecontinuity} and \eqref{eq:entropycontinuity}
\begin{equation}
n_S \vec \nabla\cdot(\vec v_N-\vec v_S)=-\frac{n^2}{s}\partial_t(s/n).
\label{eq:vnvs1}
\end{equation}
We recover $s/n=\text{const.}$ for $n_S=0$ as appropriate for the disordered phase. 
Similarly, the combination of Eq. \eqref{eq:momentumconservation} and \eqref{eq:changesuperfluidvelocity} gives
\begin{eqnarray}
\nonumber
M n_N \partial_t (\vec v_N-\vec v_S) &=& n\vec \nabla \mu -\vec \nabla p\\
&=& -s \vec \nabla T.
\label{eq:vnvs2}
\end{eqnarray}
The last equation uses the relation
\begin{equation}
\vec \nabla p=s\vec \nabla T+n \vec \nabla \mu
\end{equation}
which follows directly from the differential of $p$, Eq. \eqref{eq:differentialp}. Combining now Eqs. \eqref{eq:vnvs1} and \eqref{eq:vnvs2} yields the analogue of Eq. \eqref{eq:wave1}. 
\begin{equation}
M\partial_t^2 (s/n)=\frac{s^2}{n^2}\frac{n_S}{n_N}\Delta T.
\label{eq:wave2}
\end{equation}

One next makes an ansatz for the thermodynamic variables in the form
\begin{eqnarray}
\nonumber
p=p_0+\delta p,\quad T=T_0+\delta T\\
n=n_0+\delta n, \quad s/n=s_0/n_0+\delta(s/n),
\end{eqnarray}
where $p_0$, $T_0$, $n_0$ and $s_0$ are constant in space and time whereas $\delta p$, $\delta T$, $\delta n$, and $\delta(s/n)$ are small and vary like $\text{sin}[p(x-ct)]$. We use $\delta T$ and $\delta n$ as independent variables, with
\begin{eqnarray}
\nonumber
\delta p &=& \frac{\partial p}{\partial T}{\big |}_n \,\delta T + \frac{\partial p}{\partial n}{\big |}_T \,\delta n,\\
\delta (s/n) &=& \frac{\partial (s/n)}{\partial T}{\big |}_n \,\delta T + \frac{\partial (s/n)}{\partial n}{\big |}_T \,\delta n,
\end{eqnarray}
in order to obtain from Eqs. \eqref{eq:wave1} and \eqref{eq:wave2} the wave equation
\begin{equation}
\begin{pmatrix}Mc^2\frac{\partial(s/n)}{\partial T}{\big |}_n-\frac{s^2 n_S}{n^2 n_N} &&, && Mc^2 \frac{\partial(s/n)}{\partial n}{\big |}_T \\ -\frac{\partial p}{\partial T}{\big |}_n &&, && Mc^2-\frac{\partial p}{\partial n}{\big |}_T\end{pmatrix}\begin{pmatrix}\delta T \\ \delta n\end{pmatrix}=0.
\end{equation}
As a condition for possible sound velocities $c$ one obtains
\begin{eqnarray}
\nonumber
(M c^2)^2-(Mc^2)\left[\frac{\partial p}{\partial n}{\bigg|}_{s/n}+\frac{s^2 n_S T}{n^2 n_N c_v}\right]\\
+\frac{s^2 n_S T}{n^2 n_N c_v}\frac{\partial p}{\partial n}{\bigg |}_T=0.
\label{eq:firstandsecondsound}
\end{eqnarray}
This relation uses
\begin{equation}
c_v=T\frac{\partial(s/n)}{\partial T}{\big |}_n
\end{equation}
as well as
\begin{equation}
\frac{\partial(s/n)}{\partial n}{\big |}_T=\frac{c_v}{T}\frac{\partial T}{\partial p}{\big |}_n\left[\frac{\partial p}{\partial n}{\big |}_T-\frac{\partial p}{\partial n}{\big |}_{s/n}\right].
\end{equation}
The latter relation follows from
\begin{equation}
\frac{\partial p}{\partial n}{\big |}_{s/n} = \frac{\partial p}{\partial n}{\big |}_T + \frac{\partial p}{\partial T}{\big |}_n \frac{\partial T}{\partial n}{\big |}_{s/n}
\end{equation}
together with 
\begin{eqnarray}
\nonumber
\frac{\partial T}{\partial n}{\big |}_{s/n} &=& \frac{\partial(T,s/n)}{\partial(n,s/n)} = -\frac{\partial(T,s/n)}{\partial(T,n)}\frac{\partial(T,n)}{\partial(s/n,n)}\\
&=& - \frac{\partial(s/n)}{\partial n}{\big |}_T \frac{T}{c_v}.
\end{eqnarray}

\begin{figure}
\includegraphics[width=\linewidth]{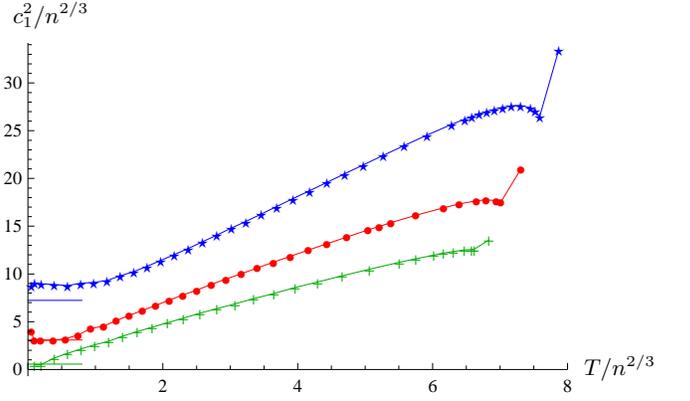}
\caption{(Color online) First velocity of sound $c_1^2/n^{2/3}$ as a function of the dimensionless temperature $T/n^{2/3}$ for the cases I (crosses), II (dots), and III (stars). We also show the prediction from Bogoliubov theory for $T\to 0$ (short solid lines).}
\label{fig:Firstvelocityofsound}
\end{figure}

\begin{figure}
\includegraphics[width=\linewidth]{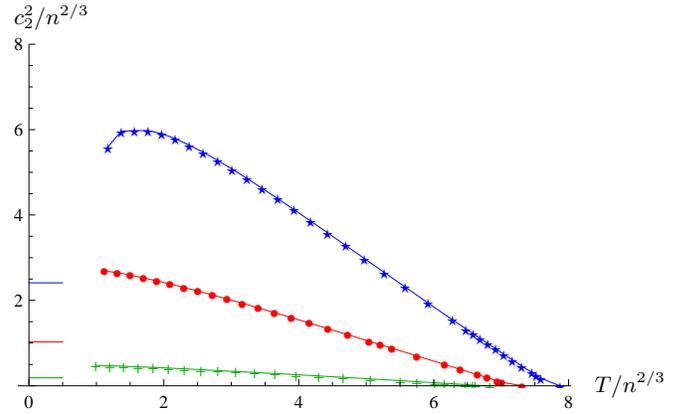}
\caption{(Color online) Second velocity of sound $c_2^2/n^{2/3}$ as a function of the dimensionless temperature $T/n^{2/3}$ for the cases I (crosses), II (dots), and III (stars). We also show the prediction from Bogoliubov theory for $T\to 0$ (short solid lines). For $T/n^{2/3}<1$ our numerical determination becomes unreliable, since $c_2^2$ is dominated by a ratio of terms that vanish for $T\to 0$.}
\label{fig:Secondvelocityofsound}
\end{figure}

With these ingredients one can now solve Eq. \eqref{eq:firstandsecondsound} for the first and second velocity of sound. The numerical results as a function of temperature are shown in Fig. \ref{fig:Firstvelocityofsound} and \ref{fig:Secondvelocityofsound}.
We also show there the prediction from Bogoliubov theory for $T\to 0$ (short solid lines). For $c_1^2$ the agreement with our findings is rather good, although there are some deviations for strong interactions as in case III. For $c_2^2$ our numerical determination becomes unreliable for $T/n^{2/3}<1$ since $c_2^2$ is dominated by the term $(s^2 n_S T)/(n^2 n_N c_v)$ in Eq. \eqref{eq:firstandsecondsound}. In the limit $T\to 0$ the quantities $s$, $n_N$, and $c_v$ also go to zero so that the numerical value for $c_2^2$ is sensitive to the precise way how this limit is approached.

We observe that Eq. \eqref{eq:firstandsecondsound} can be written as
\begin{eqnarray}
\nonumber
(M c^2)^2-\left[M v_S^2+\frac{n_S T s^2}{(n-n_S) c_v n^2}\right](Mc^2)\\
+\frac{n_S T s^2}{(n-n_S)c_v n^2}M v_T^2=0,
\end{eqnarray}
with the single fluid isothermal and adiabatic sound velocities $v_T$ and $v_S$ given by Eqs. \eqref{eq:singlefluidisothermalsound} and \eqref{eq:singlefluidadiabaticsound}. This shows that $c$ coincides with $v_S$ in the disordered phase where $n_S=0$. 
An intuitive form of the wave equation can be written as
\begin{eqnarray}
\nonumber
\partial_t^2 \delta n &=& v_T^2\Delta \delta n+ (v_S^2-v_T^2)\Delta \delta \tilde T,\\
\partial_t^2\delta \tilde T &=& (v_S^2-v_T^2+\bar v^2)\Delta\delta \tilde T + v_T^2 \Delta \delta n,
\end{eqnarray}
with
\begin{equation}
M\bar v^2 = \frac{s^2 n_S T}{n^2(n-n_S)c_v}, \quad \delta\tilde T=\frac{\delta T}{\eta}
\end{equation}
and
\begin{eqnarray}
\eta &=& -\frac{T}{c_v}\frac{\partial(s/n)}{\partial n}{\big |}_T = \frac{\partial T}{\partial n}{\big |}_{s/n}.
\end{eqnarray}
For fluctuations of $\delta n$ and $\delta \tilde T$ only $v_T$, $v_S$ and $\bar v$ matter. In the limit $T\to0$ one observes $v_S^2\to v_T^2$ such that the fluctuations $\delta n$ are governed by the isothermal sound velocity $v_T$. On the other hand, the the velocity $\bar v$ characterizes the dynamics of a linear combination of $\delta \tilde T$ and $\delta n$. 

\section{Conclusions}
For non-relativistic bosons with a repulsive pointlike interaction we have computed the dependence on temperature, density and the interaction strength of various thermodynamic observables: entropy, pressure, energy, specific heat, isothermal and adiabatic compressibility, isothermal and adiabatic sound velocity as well as the first and second velocity of sound in the superfluid phase. Non-analytic features at the critical temperature are clearly visible even though the limits of numerical differentiation do not allow a very fine resolution. The truncation of the functional flow remains rather simple, but we do not expect qualitative changes from the use of an extended truncation. The present truncation is already sufficient for reproducing correctly all expected qualitative features, both near the critical temperature of the phase transition between the superfluid and disordered phase and the quantum phase transition in the zero temperature limit. The long distance physics is under control and no infrared problems occur. Quantitative improvements may be achieved by an extension of the truncation and by increased numerical precision near the critical temperature.  

The computation of the thermodynamic response functions enters the hydrodynamic equations. They can be used for an investigation of the motions of atoms in a trap. Precise knowledge of the thermodynamics may allow for precision studies of this motion under the influence of time varying trap geometry or time varying coupling strength. 

\begin{acknowledgments}
We thank M.~K.~Oberthaler for interesting und useful discussion.
\end{acknowledgments}


\begin{thebibliography}{12}

\bibitem{FirstBEC}
M.~H.~Anderson, J.~R.~Ensher, M.~R.~Matthews, C.~E.~Wieman, and E.~A.~Cornell, Science {\bf 269}, 198 (1995);
C.~C.~Bradley, C.~A.~Sackett, J.~J.~Tollett, and R.~G.~Hulet, Phys.\ Rev.\ Lett.\ {\bf 1687} (1995);
K.~B.~Davis, M.-O.~Mewes, M.~R.~Andrews, N.~J.~van~Druten, D.~S.~Durfee, D.~M.~Kurn, and W.~Ketterle, Phys.\ Rev.\ Lett.\ {\bf 75}, 3969 (1995).

\bibitem{BoseEinstein}
A.~Einstein, Sitzungsber.\ Preuss.\ Akad.\ Wiss. 1924, 261; {\it ibid.} 1925, 3; S.~N.~Bose, Z.\ Phys.\ {\bf 26}, 178 (1924).

\bibitem{ReviewsBECExperiment}
  F.~S.~Dalfovo, L.~P.~Pitaevkii, S.~Stringari, and S.~Giorgini, Rev.\ Mod.\ Phys.\ {\bf 71}, 463 (1999);
  A.~J.~Leggett, Rev.\ Mod.\ Phys.\ {\bf 73}, 307 (2001);
	C.~Pethick and H.~Smith, {\it Bose-Einstein Condensation in Dilute Gases} (Cambridge University Press, Cambridge, 2002); 
	O.~Morsch and M.~Oberthaler, Rev.\ Mod.\ Phys.\ {\bf 78}, 179 (2006); 
	I.~Bloch, J.~Dalibard, and W.~Zwerger, {\it ibid.} {\bf 80}, 885 (2008).

\bibitem{PitaevskiiStringari}
	L.~Pitaevskii and S. Stringari, {\it Bose-Einstein Condensation} (Oxford University Press, Oxford, 2003).
	
\bibitem{Oberthalercv}
	R.~Gati, J.~Esteve, B.~Hemmerling, T.~B.~Ottenstein, J.~Appmeier, A.~Weller, and M.~K.~Oberthaler, New~J.~Phys. {\bf 8}, 189 (2006).
	
\bibitem{Donner}
	T.~Donner, S.~Ritter, T.~Bourdel, A.~\"{O}ttl, M.~K\"{o}hl, and T.~Esslinger, Science {\bf 315}, 1556 (2007).

\bibitem{BeliaevGavoretNepomnyashchii}
	S.~T.~Beliaev, Sov.\ Phys.\ JETP {\bf 7}, 289 (1958); {\bf 7}, 299 (1958);
	J.~Gavoret, P.~Nozi\`eres, Ann.\ Phys.\ (N. Y.) {\bf 28}, 349 (1964);
	A.~A.~Nepomnyashchii, Y.~A.~Nepomnyashchii, JETP Lett. {\bf 21}, 1 (1975).

\bibitem{InfraredGoldstone}
  C.~Castellani, C.~Di Castro, F.~Pistolesi, G.~C.~Strinati,
	Phys.\ Rev.\ Lett.\ {\bf 78}, 1612 (1997);
	F.~Pistolesi, C.~Castellani, C.~D.~Castro, G.~C.~Strinati,
	Phys.\ Rev.\ B {\bf 69}, 024513 (2004);
  C.~Wetterich, Phys.\ Rev.\ B {\bf 77}, 064504 (2008).

\bibitem{Wetterich:1990an}
  C.~Wetterich,
  Z.\ Phys.\  C {\bf 48}, 693 (1990);
  Nucl.\ Phys.\  B {\bf 352}, 529 (1991).

\bibitem{FunctionalRenormalization}
	K.~G.~Wilson, Phys.\ Rev.\ B {\bf 4}, 3174 (1971); 
	K.~G.~Wilson, J.~B.~Kogut, Phys.\ Rep.\ {\bf 12}, 75 (1974); 
	F.~Wegner, A.~Houghton, Phys.\ Rev.\ A {\bf 8}, 401 (1973); 
	J.~Polchinski, Nucl.\ Phys.\ B {\bf 231}, 269 (1984).

\bibitem{Wetterich:1992yh}
  C.~Wetterich,
  Phys.\ Lett.\  B {\bf 301}, 90 (1993).

\bibitem{FWBEC}
S.~Floerchinger and C.~Wetterich, Phys. Rev. A {\bf 77}, 053603 (2008).

\bibitem{Bogoliubov}
	N.~N.~Bogoliubov, Phys.~Abh.~SU, {\bf 6}, 1, (1962).

\bibitem{Pelissetto:2000ek}
  A.~Pelissetto and E.~Vicari,
  Phys.\ Rept.\  {\bf 368}, 549 (2002).


\bibitem{Landau41}
	L.~D.~Landau, J.\ Phys.\ USSR\ {\bf 5}, 71 (1941).

\end{thebibliography}
\end{document}